\documentclass[epj-spec]{svjour}
\usepackage{amsmath}
\usepackage{amsfonts}
\usepackage{amssymb}
\usepackage{graphicx}

\renewcommand{\vec}[1]{\ensuremath{\mathbf{#1}}}
\title{Comparison of models and lattice-gas simulations for Liesegang patterns}
\author{Lukas Jahnke \and Jan W. Kantelhardt\thanks{\email{jan.kantelhardt@physik.uni-halle.de}}}

\institute{Institute of Physics, Theory group, Martin-Luther-Universit\"at
Halle-Wittenberg, 06099 Halle, Germany}
\abstract{For more than a century Liesegang patterns -- self-organized,
quasi-periodic structures occurring in diffusion-limited chemical reactions
with two components -- have been attracting scientists.  The pattern formation
can be described by four basic empirical laws.  In addition to many experiments,
several models have been devised to understand the formation of the bands and
rings.  Here we review the most important models and complement them with
detailed three-dimensional lattice-gas simulations.  We show how the mean-field
predictions can be reconciled with experimental data by a redefinition of
the distances suggested by our lattice-gas simulations.}

\begin{document}
\newcommand{\erf}{\operatorname{erf}}
\newcommand{\erfc}{\operatorname{erfc}}
\renewcommand{\vec}[1]{\ensuremath{\mathbf{#1}}}
\maketitle

\section{Introduction}

In recent years the general interest in self-organized structures is growing,
triggered by the idea of cheap and fast production of nano-scaled devices.
One of the promising effects for obtaining such devices is Liesegang pattern
formation, based on a reaction-diffusion process.  Experimental evidences for
Liesegang patterns in solid materials on the nano-scale have already been
obtained, see, e.~g., \cite{Mohr2001,Bensemann2005,Grzybowski2005}.  For
example, periodic patterns of silver nano particles in glass were observed by
electron microscopy \cite{Mohr2001}. If it became possible to control the
growth of such patterns with experimentally tunable parameters, self-organized
optical devices could be made.  In addition, the Liesegang phenomenon is an
interesting research topic on its own due to the simple patterns arising out
of complicated reaction and diffusion processes.

Since the first description of Liesegang rings in gels by the German chemist
Raphael Eduard Liesegang in 1896 \cite{Liesegang1896}, many experiments and
several models have been devised to understand the formation of the bands and
rings.  Although the basic problems are solved and four universal empirical laws
have been found common to all experimentally observed Liesegang phenomena, there
are still open questions.  In literature misunderstandings regarding the exact
definition of the measured quantities cause some discrepancies between the
experimental results and the predictions of theoretical (mean-field) models.
In this paper, we review the most important models for Liesegang pattern
formation and complement them with extensive lattice-gas simulations of the
reaction-diffusion processes.  Based on our three-dimensional (3d) simulation 
results, we find a way to reconcile most experimental observations with the results 
obtained in mean-field models.  In addition, we obtain evidence that fluctuations in 3d
reaction-diffusion processes may play an important role in Liesegang pattern
formation on the nano-scale.

The paper is structured as follows.  Section~\ref{sec:basiclaws} describes the main
experimental findings, which can be summed up in four empirical universal laws
describing the Liesegang patterns.  Section~\ref{sec:models} reviews previous work on
models reproducing these laws.  After an overview including a discussion of
recent trends, we focus first on reaction-diffusion models with thresholds, and
describe afterwards the spinodal decomposition model and a kinetic Ising model.
Section~\ref{sec:latticegas} is devoted to our studies based on
lattice-gas simulations, presenting
both the numerical method, the main findings, and a suggestion for reconciling
the mean-field predictions with both, the experimental data and our 3d simulation
results by a redefinition of the distances.  Section~\ref{sec:sumoutlook} summarizes our
findings and gives an outlook on further work on Liesegang pattern formation.

\section{Experimental findings and empirical universal laws}
\label{sec:basiclaws}

In his original experiment, Liesegang covered a glass plate with a layer of
gelatin impregnated with potassium chromate \cite{Liesegang1896}. Then he
added a small drop of silver nitrate in the centre.  As a result, silver
chromate was precipitated in the form of a series of concentric rings with
regularly varying spacings.  These rings became famous
as Liesegang rings or more generally Liesegang patterns.  Shortly after the
first experiments, Wilhelm Ostwald presented an explanation for the
occurrence of the rings \cite{Ostwald1897}, which is still the basis for most
of the models today.  The next important experimental findings followed several 
years later. In 1903, Morse and Pierce \cite{Morse1903} investigated the formation 
time of the bands, observing diffusional dynamics, i.~e., the time law.  Jablczynski 
\cite{Jablczynski1923} showed twenty years later that Liesegang patterns follow 
a geometric series, i.~e., the spacing law.  Based on this observation Matalon 
and Packter \cite{Matalon1955,Packter1955} investigated the functional dependence
of the positions of the bands on the concentrations of the reacting agents in 1955.

\begin{figure}
\begin{center}\includegraphics[width=7cm]{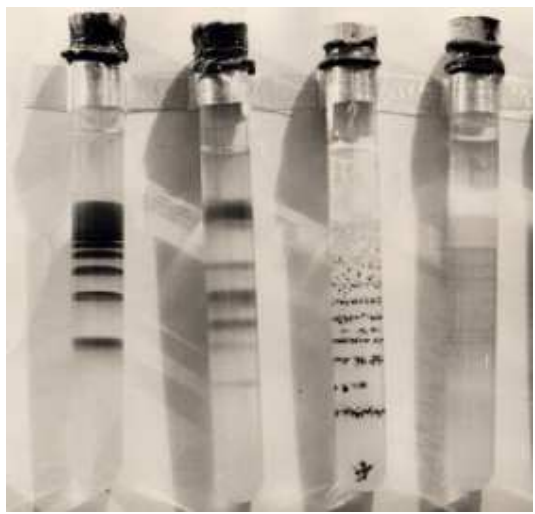}\end{center}
\caption{(colour online) Experimental examples of Liesegang patterns in test tubes. 
Agent $A$ is injected from the open end (top) of the tube, which contains agent $B$ 
dissolved in a gel, yielding a linear geometry (Figure by D.~B. Siano, 
{\tt http://commons.wikimedia.org/wiki/Image:Liesegangrings.jpg}).
\label{fig:ExpBsp}}
\end{figure}

The geometry of the pattern depends on the initial conditions for the reacting
agents.  Usually one agent (the inner electrolyte), represented by the $B$ particles 
in the models, is initially homogeneously distributed in the sample or gel.  The 
second agent (the outer electrolyte), represented by the $A$ particles, is injected.  
If $A$ is injected in the centre, precipitation rings are formed.  If $B$ is 
homogeneously distributed in a cylindrical tube and $A$ is injected from one end 
of the tube, bands form perpendicular to the motion of the reaction
front, see Fig.~\ref{fig:ExpBsp}.  The second experimental setup
is more appropriate for a theoretical description because it is effectively one
dimensional (1d) \cite{Henisch1988}.  Most of the theoretical and experimental work
was done using this linear configuration; here we also focus on the 'band' setup.
Although Liesegang rings are basically a projection of the bands onto polar
coordinates, spiral patterns have been observed in the ring configuration
exclusively.  They have no equivalent in the linear configuration.

To define the basic observables, we assume that the $n$th band forms at time
$t_n$ at distance $x_n$ from the side where $A$ is injected.  The width of the
$n$th band is denoted by $w_n$.  Firstly, the position $x_n$ is empirically
found to be proportional to the square-root of the time $t_n$,
\begin{equation} x_n \propto \sqrt{t_n}. \label{eq:timelaw} \end{equation}
This rule called {\it time law} in literature has been confirmed experimentally
many times \cite{Morse1903,Shinohar1970,Kai1982,Muller1982,Levan1987,%
Sultan2002,George2002,Muller2003,George2005}.

\begin{figure} 
\begin{center} \includegraphics[width=13.5cm]{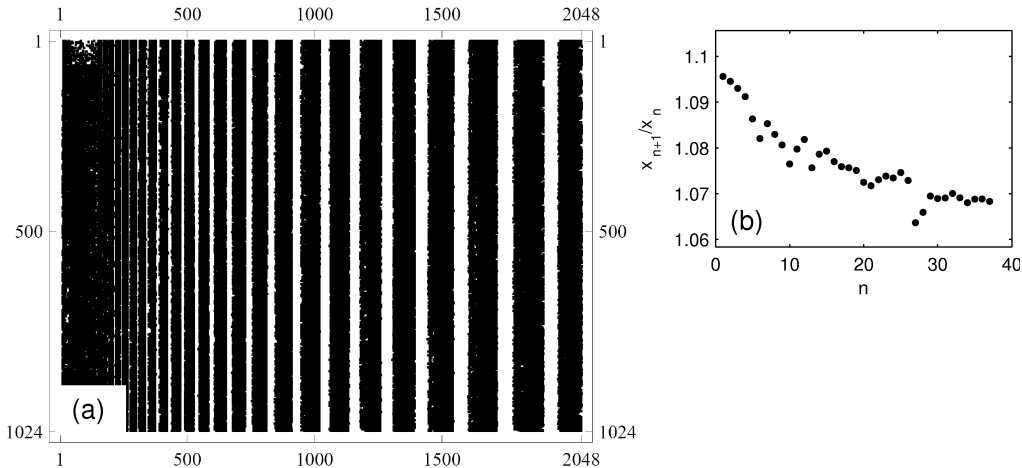}\end{center}
\caption{Results of a 3d lattice-gas simulation of Liesegang
band formation.  (a) Injecting $A$ from the left, Liesegang bands (black) with
increasing distance form in the simulated lattice of $32\times 32 \times 2048$
sites.  To obtain the projection, all 32 two-dimensional slices were placed next
to each other vertically.  (b) The ratio $x_{n+1}/x_n$ of the positions of the 
$(n+1)$st and $n$th band is plotted versus $n$. See Fig.~6 in Sect.~4.3 for the 
parameters of the simulation. \label{fig:BspLS}}
\end{figure}

Figure~\ref{fig:BspLS} shows that the position $x_n$ of the $n$th band follows
approximately a geometric series converging to the {\it spacing law}
\cite{Jablczynski1923}
\begin{equation} x_n \propto (1+p)^{n} \Leftrightarrow x_{n+1} / x_{n} \to
  1+p \quad {\rm for \; large} \; n \label{eq:spacelaw2} \end{equation}
with $p>0$ the spacing factor.  Typical empirical values for $p$ reported in
literature range from 0.05 to 0.4 \cite{Mohr2001,Bensemann2005,Muller1982,George2002,%
George2005,Narita2006}.

A first systematic experimental analysis of the functional dependence of $p$ on
the concentrations of agents $A$ and $B$ was done by Matalon and Packter
\cite{Matalon1955,Packter1955}. They gathered experimental results on $p$ and
found a functional dependence on the concentrations of both agents, $a_0$ and
$b_0$, respectively.  This dependency is known as the {\it Matalon-Packter law}
and takes the form
\begin{equation}  p=F(b_0) + G(b_0) \frac{b_0}{a_0}, \label{eq:MatalonPackterLaw}
\end{equation}
with $F$ and $G$ as dimensionless monotonously decreasing functions of $b_0$ 
\cite{foot:MPlaw}. $F$ and $G$ are not dependent on $a_0$ making $p$
linearly dependent on $1/a_0$ for constant $b_0$.  This can serve as a test for
the validity of the Matalon-Packter law.  In Sect.~\ref{sec:MPlaw} we will show
that Eq.~(\ref{eq:MatalonPackterLaw}) is just an approximation of a more general
law in the limit of reaction fronts much faster than the diffusion of the $B$
particles.  This limit holds for most of the experiments, because $a_0$ is usually
much larger then $b_0$ while both electrolytes diffuse equally fast.  An exception
seems to be the experiment with nanoscale silver particles in glass \cite{Mohr2001}
where the diffusion of silver ions, the inner electrolyte, might be one magnitude
faster than the diffusion of the outer electrolyte, hydrogen.  This leads to a
slower motion of the reaction front compared to the diffusion of the silver ions.

As can be seen in Fig.~\ref{fig:BspLS}(a) the Liesegang bands are getting broader
for larger $n$.  It is possible to set up an empirical law describing the width
$w_n$ of the $n$th band as function of the position $x_n$.  Experimentalists use
two competing versions of the {\it width law},
\begin{equation}  w_n = \mu_1 x_n + \mu_2 \qquad {\rm and} \qquad w_n \propto
x^{\alpha}_n \label{eq:widthlaw} \end{equation}
see, e.~g., \cite{Kai1982,Narita2006,Dee1986} and \cite{George2005,Chopard1994,Racz1999,Droz1999}, 
respectively.  We are not aware of any papers comparing the two versions.  In addition, 
different values of the exponent $\alpha$ in the second version have been published.
The experiment with nanoscale silver particles in glass can be fitted by $\alpha \approx 
0.7$ \cite{Mohr2001} (using the second version of Eq.~(\ref{eq:widthlaw})), but the results 
can be equally well fitted by the first version.  Most other papers report larger values
of $\alpha$, e.~g., $0.9 < \alpha < 1$ \cite{Droz1999}.  Publications of early reactive 
lattice-gas simulations, where the second version was first introduced, fit best with 
values of $\alpha \approx 0.5-0.6$ \cite{Chopard1994,Chopard1994b}, but no comparison with 
the first version of Eq.~(\ref{eq:widthlaw}) was done.  It is difficult to distinguish 
between both versions of the width law because the number of bands is limited and the widths 
$w_n$ of the bands are small compared to their positions $x_n$.  Theoretical works prefer 
the second version with $\alpha=1$ \cite{George2005,Racz1999,Droz1999}. In 
Sect.~\ref{sec:reconwidth} the different versions will be tested on our results of 
lattice-gas simulations. We will also propose an alternative, theoretically well grounded 
approach unifying both versions there.

The time law (\ref{eq:timelaw}) is a simple consequence of a diffusion process.
The position $x_f(t)$ of the reaction front between $A$ and $B$ moves proportional
to $\sqrt{t}$ with a prefactor that depends on $a_0$ and $b_0$ \cite{Crank1996}.
The other three laws cannot be explained so easily; one needs models describing
nucleation and growth of the bands.  These models can be categorized into two types.
The first type is based on diffusion and reaction dynamics plus some thresholds which
account for nucleation and growth.  There exists a wide range of modifications.  The
second type of models uses well established phase separation techniques to explain
the pattern formation based on spinodal decompositions \cite{Antal1999} or a kinetic
Ising model \cite{Antal2001}.  All of these models will be reviewed in the next section.

\section{Models} \label{sec:models}

\subsection{Overview and recent trends}

Reaction-diffusion models with thresholds are the oldest models describing
Liesegang pattern formation.  Only a few month after the first experiments by
Liesegang, Ostwald gave a possible explanation of the pattern formation process
on the basis of supersaturating liquids \cite{Ostwald1897}.  He suggested that
the precipitation is not a result of a balanced reaction but must happen
spontaneously when the concentration product $K$ of the reactive partners reaches
a critical concentration $K_{sp}$.  The precipitation grows until $K$ falls under
a second concentration $K_p$.  Both thresholds set the stage for several models
differing in the detailed description of the reaction process.  In principle
they all follow a reaction scenario
\begin{equation}  A + B \to \cdots C \cdots \to D, \label{rec:ABCD}
\end{equation}
and differ in the way the intermediate stage $\cdots C \cdots$ and the precipitation
$\to D$ are described.  For the concentrations of the different chemical
agents $A,B,\ldots$ the symbols $a,b,\ldots$ will be used in the following.

The empirical laws described in the previous section became the basis for the
theoretical work which started in the fifties.  First analytical descriptions
\cite{Wagner1950,Prager1956,Zeldovich1961} showed that the patterns can be
explained by diffusion and reaction processes with a moving reaction front.  When
numerical simulation became feasible about 90 years after the first experiments
it was possible to reproduce patterns dynamically by studying the reaction-diffusion
equations \cite{Levan1987,Dee1986}.  The fact that a novel approach was introduced
in 1999 \cite{Racz1999,Antal1999} shows that the Liesegang pattern phenomena are
still an active research field in both, experiment and theory.

Apart from the novel simulation approach the trends in the literature follow two 
general lines.  The first line works on an analytical description of the basic laws
\cite{Droz1999,Antal1998,Lebedeva2004} or even looks for more general laws
\cite{Izsak2005}.  The second line varies the initial conditions and the
geometrical configurations of the setup to understand the phenomena better and
to find ways for applying the pattern formation in interesting engineering
problems.  Alternative geometries \cite{Izsak2004} and complex 3d
boundary conditions \cite{Bensemann2005,Grzybowski2005,Smoukov2005} have been
studied experimentally.  Variation of the reaction terms \cite{Chopard1999},
additional terms for a dissolution of the bands \cite{Sultan2002,Msharrafieh2006},
as well as open systems \cite{Lagzi2005} and, last but not least, systems with an
additional electric field \cite{Sultan2002,Lagzi2003,Shreif2004,Bena2005b}
have been studied to vary the patterns.  In very recent work the motion of the
reaction front is even detached from the phase separation, which might be initiated,
e.~g., by a temperature gradient \cite{Antal2007}.  One hopes to understand and
control the pattern formation process such that bands with constant distance can be
generated, see \cite{Bensemann2005,Grzybowski2005} for experimental work in this
direction.

Most of the theoretical work is based on an analytical study or numerical
solution of differential equations.  Since such equations are always based on the
concentrations $a,b,\ldots$ of the reacting agents, they generally yield mean-field
solutions.  Although such mean-field solutions can reproduce the basic laws listed
in the previous section they have two disadvantages.  Firstly, they cannot account
for the statistical character of the reactions.  Hence, the influence of particle
number fluctuations and thus the stability of the patterns cannot be investigated
adequately.  Although there are some ideas to include fluctuation by an additional
noise term in mean-field models \cite{Bensemann2005,Lagzi2003}, we think that
models with intrinsically statistical character are more adequate.  This is
particularly true for mesoscopic and nanoscopic systems, where fluctuations become
more important.  First experiments on microscopic scales indicate that fluctuations
might play an important role \cite{Mohr2001,Bensemann2005,Grzybowski2005}.
Secondly, the differential equations may not describe the microscopical structure
of the bands.  Chopard {\it et~al.} \cite{Chopard1994,Chopard1994b} proposed an
alternative approach by taking the basic principles of the mean-field description
and implementing them in a reactive lattice-gas simulation.  Such simulations can
be very helpful in reconciling mean-field predictions with experimental
data, as we will see in Sect.~\ref{sec:latticegas}.  A similar approach seems also 
possible using Ising models \cite{Antal2001}.

In the following sections, the mentioned quantitative models and simulations will
be reviewed, except for the lattice-gas simulations to be presented in 
Sect.~\ref{sec:latticegas}.

\subsection{Ion product saturation models}
\label{sec:Ionproduct}

In the first and easiest quantitative models for Liesegang pattern formation, the
ion product models, the precipitation takes place without an intermediate stage;
i.~e., there is no $C$ stage in (\ref{rec:ABCD}) \cite{Wagner1950,Prager1956,%
Zeldovich1961}.  Like in the Ostwald model, nucleation $A+B \to D$ occurs when
the local product concentration $K=ab$ reaches the threshold $K_{sp}$.  The
precipitates of $D$ will grow and deplete their surrounding of $A$ and $B$.  As
the reaction front proceeds, the product concentration around the immobile
precipitate decays until growth becomes impossible.  Wagner \cite{Wagner1950},
Prager \cite{Prager1956} and Zeldovich {\it et~al}. \cite{Zeldovich1961} could
show that these ingredients yield patterns
which obey the time law (\ref{eq:timelaw}) and spacing law (\ref{eq:spacelaw2}).
The diffusion profiles of $a$ and $b$ are described by a system of coupled
integro-differential equations with given boundary conditions.  These equations
can be rewritten in a more convenient way, if local coordinates are introduced
in the form $\lambda_n = x_n/(2 \sqrt{D t_n})$ and $\gamma_n = x_n/x_{n-1}$.
Taking $n \to \infty$ and $x_0 \to 0$ renders the equations
mathematically solvable.  The local coordinates $\lambda_n
\to \lambda$ and $\gamma_n \to \gamma$ yield the time law and the spacing law, respectively.
However, since these laws were the result of a continuous limit, the dynamics of
the process was lost.  Furthermore, the formation of the Liesegang bands was taken
for granted and not proved by solving reaction-diffusion equations.

The first dynamical version of the ion product model proposed by Ross {\it et~al.}
\cite{Levan1987} was deduced from the elementary chemical reaction $A+B \to D$
and solved numerically.  Besides the typical diffusion terms for $A$ and $B$,
\begin{eqnarray}
  \partial_t a &=& D_A \partial^2_x a - R \quad {\rm and}  \label{eq:ab1} \\
  \partial_t b &=& D_B \partial^2_x b - R \label{eq:ab2}
\end{eqnarray}
with diffusivities $D_A$ and $D_B$, respectively, the equations include a reaction
rate $R$ constructed for a specific experimental configuration.  In principle, $R$
mimics an auto-catalytic growth process with a growth rate proportional to an
increasing function of the supersaturation $S=(a b /K_{sp})-1$. The authors tested
different power laws and exponential functions and
observed pattern formation for a very quickly growing function $R(S)$ only.  Beside the
confirmation of the time law no quantitative observations of the other laws were reported.

Later Zrinyi {\it et~al}. \cite{Buki1995} pointed out that no detailed description of 
the agent transport is needed, since precipitation takes place on a faster time scale.
It is thus possible to model the growth process by a given critical threshold
$K_p$.  The corresponding source term for additional $D$ can be written as
a Heaviside step function, $c_1(a,b,d) \, \Theta(ab-K_p)$.  The growth starts when
the product concentration exceeds a second threshold $K_{sp} > K_p$, where the
rate is independent of $d$.  The partial differential equation for the precipitate
concentration $d$ can thus be written as \cite{Buki1995,foot:Zrinyi}
\begin{equation} R = \partial_t d = c_1(a,b,d) \, \Theta(ab - K_p) + c_2
\Theta(ab - K_{sp}),  \label{eq:dd} \end{equation}
where $c_1$ and $c_2$ are model-dependent functions of the parameters (initial
concentrations and diffusion coefficients), and $c_1$ also depends on $a$, $b$ and $d$.  
Liesegang patterns observed in a wide range of systems can be modelled this way.  In 
particular, it was possible to show that the time law and the spacing law hold for a 
large set of parameters \cite{Buki1995}. The Matalon-Packter law
was confirmed only if the initial concentration of the outer electrolyte $a_0$
is sufficiently large.  However, $p$ was shown to stay a monotonously decreasing
function of $a_0$ even if $a_0$ is small.  Furthermore, $p$ increases monotonously
with $K_{sp}$ in a non-linear way, and it is anti-proportional to the diffusivity 
$D_A$ of the outer electrolyte.  The results for a specific $p$ do not
depend on the individual parameters but on ratios of parameters with equal dimension
-- a simple consequence of the mean-field character of Eqs.~(\ref{eq:ab1}) to
(\ref{eq:dd}).  The findings are consistent with an experimental categorisation of
the patterns by the product and the difference of the initial concentrations
\cite{Muller1982}.  The width law, however, was not investigated.

In closely related approaches, growth was modelled proportional to the concentration
of the precipitate, corresponding to $K_p=0$ and $c_1 \propto d$ in Eq.~(\ref{eq:dd})
\cite{Antal1998,Lebedeva2004}.  Focusing on the
Matalon-Packter law (\ref{eq:MatalonPackterLaw}), Antal {\it et~al.} found that $F$ is
constant and $G \propto \sqrt{D_B/D_A} \, K_{sp}/b^2_0$ \cite{Antal1998}.
This result was achieved analytically based on the assumption $b_0 \ll a_0$; hence
the failure of the Matalon-Packter law for low $a_0$ could not be observed.  This result is
inconsistent with $p \propto 1/D_A$ in \cite{Buki1995}, due to different assumptions for
$c_1$ and $c_2$.  Lebedeva {\it et~al.} searched for a criterion for the pattern
formation \cite{Lebedeva2004}, finding $\Phi=K_{sp} k_2 D_B/u^2(t) > 2+\sqrt{5}$,
with $k_2$ the growth rate and $u(t)$ the velocity of the front (decreasing with
time).  Patterns are thus formed when the nucleation threshold $K_{sp}$ is high
and the growth rate $k_2$ is large.  The time dependence of $u(t)$ explains why
patterns do not start at $x_0=0$ but an empty (``plug'') zone exists at the edge.
The first two results match with those of Ross {\it et~al.} \cite{Levan1987} and Zrinyi
{\it et~al.} \cite{Buki1995}, who did not study the velocity of the reaction front.

\subsection{Nucleation and growth models}\label{sec:nucleationgrowth}

The next models with slightly increased complexity separate the
reaction-diffusion process $A+B \to C$ from the nucleation and growth process
$C \to D$, introducing an intermediate state $C$, see Eq.~(\ref{rec:ABCD}).  This,
however, considerably simplifies the analysis.  In the literature such models are
often called nucleation and growth models because the formation of $D$ depends
on the nucleation and accumulation of $C$ rather than on the product
concentration of $A$ and $B$.

This model was first introduced by Keller {\it et~al.} \cite{Keller1981} who
analysed reaction-diffusion equations and confirmed the time law
(\ref{eq:timelaw}) and the spacing law (\ref{eq:spacelaw2}).  In contrast to
Wagner and Prager \cite{Wagner1950,Prager1956} they could calculate the
positions $x_n$ of the bands without a priori assuming band formation and
stopping of the band growth.  The first numerical solutions for a nucleation
and growth model were presented by Dee \cite{Dee1986}.  He used a similar
technique as Ross {\it et~al.} \cite{Levan1987} (see previous section) and
determined the nucleation and growth criterion by classical nucleation
theory.  A detailed work concerning the nucleation and growth of silver
particles similar to Dee was presented recently \cite{Kaganovskii2007}.
Since the reaction-diffusion process can be separated from the
precipitation process, we will discuss the two stages separately.

\subsubsection{First process: reaction-diffusion $A+B \to C$}\label{sec:ABC}

\begin{figure}
\begin{center} \includegraphics[width=13cm]{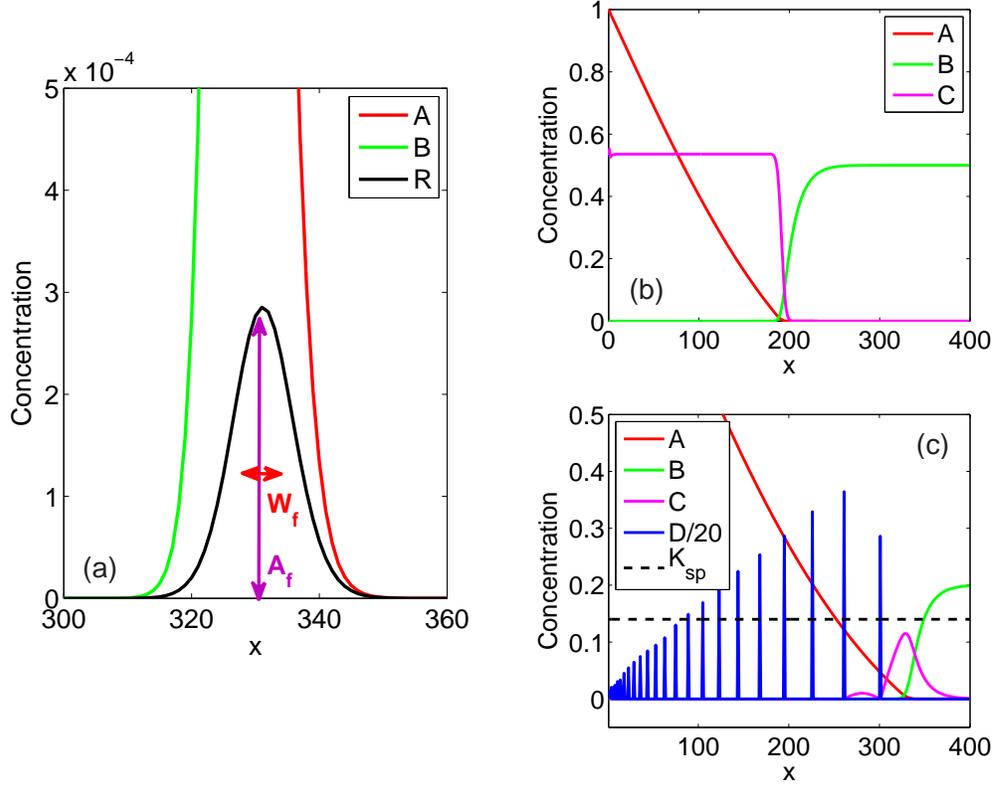}\end{center}
\caption{(colour online) Illustration of the concentration profiles obtained 
in the reaction-diffusion process $A+B \to C$ without (a,b) and with (c) 
precipitation $C \to D$.  The concentrations $a(x,t)$ (red) and $b(x,t)$ 
(green) are shown for fixed time $t$ together with (a) the reaction rate
$R(x,t) \propto a b$ (black) following Eq.~(\ref{eq:ReakScale}), (b) the 
accumulated concentration $c(x,t)$ (magenta), and (c) the rescaled concentration 
of the precipitate $D$ (blue).  Note the constant value of $c(x,t) = c_0$ 
behind the reaction front in (b) according to Eq.~(\ref{eq:c0}). The dashed 
line in (c) indicates the threshold $K_{sp}$; the next band is started when
$c(x,t)$ reaches $K_{sp}$. \label{fig:nucgrowth}} \end{figure}

The simple reaction-diffusion process $A+B \to C$ is important
independently of Liesegang patterns since it is a basic process in many chemical
reactions.  To apply the results to the Liesegang pattern phenomena we will
focus on a quasi-1d geometry where the reacting particles are
separated at time $t_0=0$ with initial concentration $a_0$ and $b_0$.  The
concentration profiles $a(x,t)$ and $b(x,t)$ for $x \ge 0$ and $t>0$ are
determined by the reaction-diffusion equations (\ref{eq:ab1}) and (\ref{eq:ab2})
with the initial conditions $a(x,0)=a_0 \Theta(-x)$, $b(x,0)=b_0 \Theta(x)$ 
\cite{foot:a0}. The reaction term $R$ is again assumed to be proportional to 
the concentration product $K=ab$. Under these assumptions it is possible
to calculate $a(x,t)$ and $b(x,t)$ as well as the reaction rate $R(x,t) \propto
ab$ asymptotically for large times $t$ \cite{Galfi1988},
\begin{equation} R(x,t) \sim A_f \; S_R\!\left[\frac{x - x_f(t)}{w_f(t)}
\right], \label{eq:ReakScale} \end{equation}
where $S_R$ is approximately a Gaussian \cite{Antal1999,Larralde1992}.
The result is illustrated in Fig.~\ref{fig:nucgrowth}(a).
$R(x,t)$ describes the reaction front and reaches its maximum value at $x_f$
which scales as $x_f(t) \sim \sqrt{t}$.  The width of the front scales as
$w_f(t) \sim t^{\gamma}$ with $\gamma=1/6$.  The production rate $A_f$ of
$C$ at $x=x_f$ is proportional to $t^{-\zeta}$ with $\zeta=2/3$.
A scaling analysis regarding the impact
of fluctuations \cite{Cornell1991} leads to the conclusion that the critical
dimension, above which fluctuations become unimportant, is $d_c=2$.  In 2d 
logarithmic corrections apply and in 1d the whole
dynamics change dramatically \cite{Cornell1991}.

It is possible to approximate the effective diffusion coefficient $D_f$ for the
reaction front, i.~e., the prefactor in $x_f = \sqrt{2 D_f t}$
as well as
the concentration $c_0$ of $C$ behind the reaction front.  This requires
a few assumptions \cite{Koza1996} which will be usually fulfilled in typical
experimental setups.  The reaction front $R(x,t)$ has to be confined in a
region (reaction zone) $x_f(t) - w_f(t)/2 < x < x_f(t) + w_f(t)/2 $ for all
times $t$, disregarding all reactions outside this region.  Then the
concentration profile of each agent can be written as a solution of a
diffusion equation \cite{Crank1996}.  Furthermore, Eqs.~(\ref{eq:ab1}) and
(\ref{eq:ab2}) must be approximated by a quasi-stationary solution in the
region $-(D_A t)^{1/2} \ll x \ll (D_B t)^{1/2}$.  Under these conditions the
effective diffusion coefficient $D_f$ is given by an analytic implicit equation
\cite{Koza1996},
\begin{equation}  H\!\left(- \sqrt{\frac{D_f}{2 D_A}}\right) = \frac{a_0
\sqrt{D_A}}{b_0 \sqrt{D_B}} \; H\!\left(\sqrt{\frac{D_f}{2 D_B}}\right)
\label{eq:Dfeins}\end{equation}
with $H(x)=[1-\erf(x)]\exp(x^2)$. Here, $\erf(x)$ is the Gaussian error
function defined as $\erf(x)= \frac{2}{\sqrt{\pi}} \int_0^x \exp\left(-
z^2 \right) dz$.  Evidently, the velocity of the reaction front does
not depend on each of the initial concentrations, but on their ratio
$a_0/b_0$, and it scales in the same way as $D_A$ and $D_B$, since no
characteristic time and length scales exist.  The same approximation also
yields that the concentration of $C$ behind the reaction front is constant 
(see Fig.~\ref{fig:nucgrowth}(b)), given by \cite{Antal1998}
\begin{equation} c_0 = \frac{b_0}{\pi} \sqrt{\frac{2 D_B}{D_f}} \;
H^{-1}\!\!\left(\sqrt{\frac{D_f}{2 D_B}}\right). \label{eq:c0} \end{equation}
The interesting point of these results, for the Liesegang pattern phenomena,
is that only $c_0$ and $D_f$ are essential parameters.  Since both depend
only on the initial concentrations and the diffusion coefficients, the
band formation is independent of the reaction rate which is hard to measure.
This does not imply, however, that the width of the bands or the specific
concentration profiles are independent of the reaction rate.

\subsubsection{Second process: precipitation $C \to D$} \label{sec:CtoD}

Describing the $A+B \to C$ process exclusively by $c_0$, $D_f$, and the
Gaussian reaction front (\ref{eq:ReakScale}), we now focus on the
precipitation step.  Like Ross {\it et~al.} \cite{Levan1987} 
(see Sect.~\ref{sec:Ionproduct}),
Dee \cite{Dee1986} modelled the precipitation process using the classical
droplet theory.  Droz {\it et~al.} \cite{Droz1999} showed that Dee's approach
can be simplified into 
\begin{equation} \partial_t d(x,t) = c_1 N[c(x,t),d(x,t)] + c_2
\Theta[c(x,t) - K_{sp}] \qquad {\rm with} \quad N(c,d) \propto d(x,t),
\label{eq:ualternative} \end{equation}
corresponding to Eq.~(\ref{eq:dd}) with $K_p=0$ and $c_1$ and $c_2$ constants
or proportional to $c$.  A typical result of such a simulation is shwon in 
Fig.~\ref{fig:nucgrowth}(c). Using this model, Dee
confirmed the spacing law (\ref{eq:spacelaw2}) and found the linear form
of the width law (\ref{eq:widthlaw}).  Due to limited computational
resources he was restricted to one example with only six bands and
obtained no information on the Matalon-Packter law (\ref{eq:MatalonPackterLaw}).

In Eq.~(\ref{eq:ualternative}) the growth of the precipitates is restricted
to the points of the nucleated particles, see also Fig.~\ref{fig:nucgrowth}(c). 
To obtain bands with a macroscopic width $w_n$ one would need nucleation at every 
point in $w_n$ since no non-local term exist.  Although no macroscopic width 
is reproduced in this model one sees that the hight of the bands, which can be 
interpreted as their mass or integrated particle number, seems to grow linearly. 
This is in agreement with a theoretical prediction which will be introduced in 
Sect.~\ref{sec:reconwidth}.

Alternatively it would be possible to allow non-local growth
such that $C$ can become $D$ in a broader surrounding of an initial nucleation
centre.  A popular version of the non-local term is \cite{Bensemann2005,%
Fialkowski2005}
\begin{equation} N(c,d) = \Theta\!\left[ \int_{x-dx}^{x+dx} d(x',t) \, dx'
\right] \Theta[c(x,t)-K_p]. \label{eq:NDC} \end{equation}
In this case, a second threshold $K_p$ is needed to stop the growth.  This
threshold is not needed in Eq.~(\ref{eq:ualternative}) because local growth
stops if nucleation stops.

\subsubsection{Quantitative mean-field predictions for $p$ and the
Matalon-Packter law} \label{sec:MPlaw}

It is possible to calculate the functional dependence of $p$ in the nucleation
and growth model described above \cite{Antal1998},
\begin{equation} p = \frac{D_C}{D_f} \, \left[\frac{c_0}{K_{sp}} - 1 -
\frac{D_C}{2 D_f}\right]^{-1}.  \label{eq:pMF} \end{equation}
This surprisingly simple analytical solution determines $p$ from the basic
parameters $D_f$ and $c_0$ (from $A+B \to C$), $D_C$ (for the diffusion of
$C$), and the nucleation threshold $K_{sp}$.  As expected for a mean-field
solution, Eq.~(\ref{eq:pMF}) depends only on dimensionless ratios of the
parameters.  The first term characterizes the velocity of $C$ versus the
velocity of the reaction front, representing a measure of how fast $c_0$
could be reached.  If $D_f \ll D_C$, $C$ will quickly leave the reaction
front and diffuse away.  In this case it takes longer to reach the desired
threshold $K_{sp}$, and $p$ becomes larger.  The second term makes sure
that no precipitation occurs if $K_{sp} > c_0$.  If $K_{sp}$ is close to
$c_0$ it takes longer to reach $K_{sp}$ and $p$ is also large.

The structure of Eq.~(\ref{eq:pMF}) will be similar if nucleation and growth
is modelled somewhat differently.  All mean-field solutions must scale, and
patterns emerge only if $K_{sp} \le c_0$.  For this reason Eq.~(\ref{eq:pMF}) can
serve as reference to test further properties of Liesegang pattern formation.
One of these properties is the Matalon-Packter law, which is clearly
inconsistent with Eq.~(\ref{eq:pMF}), since it does not scale linearly with
$1/a_0$.  However, Antal {\it et~al.} \cite{Antal1998} could show that the 
Matalon-Packter law is a special
case of Eq.~(\ref{eq:pMF}), if $b_0 \ll a_0$.  Then $D_B \ll D_f$ and the
Gaussian error function can be approximated by $\erf(x) \approx 1 - \exp(-x^2)
\, \left(1-\frac{1}{2 x^2} -\ldots \right)/\sqrt{\pi x}$ \cite{Abramowitz1972}.
Applying this to Eq.~(\ref{eq:c0}) we obtain $c_0 \approx
b_0 \, (1 + \frac{D_B}{D_f})=b_0 \, (1 + \frac{D_B}{D_A} \frac{D_A}{D_f})$.
In the range $0.05 \le b_0/a_0 \le 0.1$ interesting for experiments, Antal {\it et~al.}
showed numerically that $D_A/D_f$ is linear in $a_0/b_0$.  This leads to
$D_A/D_f \approx \nu_1 + \nu_2 b_0/a_0$, yielding an approximation for $c_0$,
\begin{equation} c_0 = b_0 \left(1 + \nu_1 \frac{D_B}{D_A} + \nu_2
\frac{D_B}{D_A} \frac{b_0}{a_0} \right), \label{eq:c02} \end{equation}
with $1+\nu_1 \frac{D_B}{D_A}$ and $\nu_2 \frac{D_B}{D_A}$ numbers of order
one.  Since $b_0/a_0 \ll 1$ it is possible to neglect the last term in
Eq.~(\ref{eq:c02}).  It is not possible to argue this way already in the
linear approximation of $D_A/D_f$ because the $\nu$s do not have the same
magnitude.  Inserting the approximation back into Eq.~(\ref{eq:pMF}) we obtain
\begin{equation}  p = \frac{D_C K_{sp} \nu_1}{D_A (\sigma_1 b_0 - K_{sp}) }
+\frac{D_C K_{sp} \nu_2 b_0}{D_A (\sigma_1 - K_{sp})}\frac{b_0}{a_0 }
  = F(b_0) + G(b_0)\frac{b_0}{a_0}, \label{eq:MatalonPackAntal} \end{equation}
which is in the form of Eq.~(\ref{eq:MatalonPackterLaw}).  The assumptions
needed to derive Eq.~(\ref{eq:MatalonPackAntal}) will be correct for most
macroscopic experiments.  An important exception might be given by an experiment
of nanoscale particles in glasses \cite{Mohr2001}.  In that case $a_0$ and
$b_0$ have the same magnitude, but $D_B$ seems to be one order of magnitude larger 
than $D_A$.  This leads to $c_0 \gg b_0$ in contradiction with Eq.~(\ref{eq:c02}).
However, since only one experiment was carried out yet, we cannot see whether
Eq.~(\ref{eq:pMF}) holds.

\subsubsection{The width law}\label{sec:widthlaw}

The width law has been neglected in the discussion of Liesegang pattern formation
for a long time because of difficulties in defining and measuring the width $w_n$
of the bands consistently.  To obtain a finite width in
microscopical models one needs to introduce a non-local growth
term and a second threshold as $N(c,d)$ from Eq.~(\ref{eq:NDC}).
On the other hand it is also possible to generate a width using
Eq.~(\ref{eq:ualternative}) for macroscopic models, e.~g.~\cite{Dee1986}.

The first theoretical step in the direction of clarifying the law was done by
Droz {\it et~al.} \cite{Droz1999} using a model based on Eq.~(\ref{eq:ualternative}).
They showed that the total mass $m_n$ of $D$ material in the $n$th band is
proportional to $x_n$.  The concentration profile $d(x,t)$
for $t \gg t_{n+1}$ will be stationary and has a scaling function of the form,
$d_n(x,t) = A_n d[(x - x_n)/w_n]$ for $x_n \le x \le x_n + w_n$, where the
amplitude $A_n$ could depend on $n$.  It is thus possible to relate $m_n$ with
$d_n$ by integration, $m_n = A_n \int_{x_n}^{x_n + w_n} d[(x - x_n)/w_n] \, dx =
\gamma A_n w_n \propto x_n$, where $\gamma$ is the substituted integral over
$d_n$.  Using the first version of Eq.~(\ref{eq:widthlaw}), $w_n \sim x_n^\alpha$,
this yields $A_n \sim x_n^{\beta}$ with $\alpha + \beta = 1$ \cite{Droz1999}.
The authors also discussed experiments, fitting the results with $0.9<\alpha<1$.

A more direct treatment was proposed by Racz \cite{Racz1999}. Because the $A+B \to
C$ process is independent of the $C \to D$ process the number of $C$ particles
produced in the first process is the same as the number of $D$ particles ending up in 
the bands.  The number of $C$ particles can than be calculated using the assumption that
the constant concentration $c_0$ behind the front is really reached.  Later, $C$
precipitates into bands of $D$ with a high concentration $d_h$ while the low
concentration $c_l$ of $C$ remains between the bands.  The particle conservation
law yields $(x_{n+1} - x_n) c_0 = (x_{n+1}-x_n-w_n)c_l + w_n d_h$ leading to the
width law in the form
\begin{equation} w_n = p \frac{c_0 - c_l}{d_h - c_l} x_n \approx p \frac{c_0}{d_h}
x_n, \label{eq:wn} \end{equation}
since $c_l$ is usually very small.  If $d_h$ varies with the band position as
$d_h \sim x_n^{\beta}$ the width law takes again the form $w_n \sim x_n^{\alpha}$
with $\alpha+\beta=1$.

\subsection{Further reaction-diffusion models and comparison}

For charged $C$ particles it is possible to extend the nucleation and growth model
such that nucleation only
occurs if the concentration of the outer electrolyte $A$ exceeds a third threshold
$K_a$.  Such models are known as induced sol coagulation models
\cite{Shinohar1970,Dahr1922,Dahr1925}.  The additional threshold is motivated by the fact
that the repulsive electrostatic interaction between the $C$ ions can be screened
by $A$ particles.  The model yields band formation
significantly behind the reaction front as observed in some experiments
\cite{Kai1982}.  The functional form of $p$ deviates from Eq.~(\ref{eq:pMF})
\cite{Antal1998},
\begin{equation} p = 2 \frac{D_C}{D_f} \, \left\{ \left[\frac{c_0}{K_{sp}}
\left( 1-K_a/a_0 \right)\right]^2 - \frac{D_C}{D_f} \right\}^{-1}.
\label{eq:pSG} \end{equation}

A fourth category of models is called competitive growth models
\cite{George2002,Flicker1974,Feeney1983,Venzl1986,Chernavskii1991,Chacron1999,Krug1999}.
In these models, the precipitates $D$ can dissolve with a probability decreasing
with increasing size of the precipitates.  In special cases it is possible that
the bands move \cite{Zrinyi1991}, dissolve and reprecipitate \cite{Sultan2002},
such that the total number of bands leads to a chaotic time series.

It is possible to compare the $F$ and $G$ functions of the Matalon-Packter
law (\ref{eq:MatalonPackterLaw}) for the supersaturation models with thresholds
discussed in Sects.~\ref{sec:Ionproduct} to
\ref{sec:nucleationgrowth} and Eq.~(\ref{eq:pSG}).  As already mentioned, the law is
valid only for $b_0/a_0 \ll 1$ and $D_B / D_f \le 1$.  The ion product
supersaturation model (see Sect.~\ref{sec:Ionproduct}) predicts $F(b_0) \sim const$ 
and $G(b_0) \sim K_{sp}/b^2_0$, which is in contradiction to experimental results
yielding monotonously decreasing functions of $b_0$ for $F$ and $G$
\cite{Matalon1955}.  The nucleation and growth model predicts $F(b_0) \sim
G(b_0) \sim K_{sp}/(\sigma b_0 - K_{sp}) \approx K_{sp}/b_0$ if $K_{sp} \ll b_0$.
The induced sol coagulation model predicts $F(b_0) \sim K_{sp}^2/b_0^2$ and
$G(b_0) \sim (\alpha/b_0^3 + \beta/b_0^2)$.  In addition, Antal {\it et~al.} 
\cite{Antal1998} remark that a refined version of the sol coagulation
model will converge to the nucleation and growth model in the limit $K_a/a_0
\to 0$.  With such a model it should be possible to vary the functional
dependence of $F$ such that $F(b_0) \sim b_0^{-\gamma}$ with $ 1 \le \gamma
\le 2$, corresponding to $0.2 \le \gamma \le 2.7$ observed in experiments, see
\cite{Antal1998} and references therein. Such refinement is only possible if the $C$
particle are charged. For uncharged $C$ particles the nucleation and growth model still
yields the best predictions.

Comparing all four supersaturation models one can conclude that the nucleation
and growth model serves best as a reference model.  It is the easiest model
yielding the Matalon-Packter law.  Although it needs an intermediate state $C$,
the model is actually most suitable for analytical and numerical studies.
The sol coagulation model and the competitive growth models, on the other
hand, can explain details observed in specific experimental setups, but they
are not needed to explain the basic universal laws discussed in 
Sect.~\ref{sec:basiclaws}.  The variety of models yielding these laws is remarkable.
The same is true for the variety of physical, chemical and even geological
systems showing the phenomenon.  This fact suggests
that a very general mechanism governs the pattern formation.  Such a mechanism
will be described in the next section.

\subsection{The spinodal decomposition model with Cahn-Hilliard dynamics}
\label{sec:spinodal}

Although the supersaturation models described in the previous sections
account for most of the Liesegang phenomena, there are drawbacks.
Firstly, the threshold parameters controlling the growth of the bands are
difficult to grasp theoretically and not easy to control experimentally.
Secondly, it is difficult to derive how the band formation could be
manipulated in a desired way, since the structure of the models is too
complicated.  Furthermore, the specific models might seem insufficiently
universal.  Hence, a new approach free of thresholds and reaction-diffusion
equations was recently suggested by Antal {\it et~al.} \cite{Antal1999}.
They studied only the second process $C \to D$ of Eq.~(\ref{rec:ABCD})
applying the spinodal decomposition theory for phase separating processes
\cite{Gunton1983,Domb1983} to describe the phase separation into bands.
We note that the distinction between $C$ and $D$ is actually not necessary
here, since the bands correspond to areas with high concentration $c=d_h$,
while there is little $C$ between the bands, $c=c_l$ (see also
Sect.~\ref{sec:widthlaw}). Although the model was introduced to describe 
$C \to D$ processes it is equally valid for processes where the $C$ 
particles arrange in bands of high and low concentrations. Then the phase 
separation is assumed to take place at a very low effective temperature 
leading to stable bands after long times.

As we have seen in the Sect.~\ref{sec:ABC} the production of $C$ particles
in the $A+B \to C$ reaction is well understood.  The reaction front --
an approximately Gaussian shaped region where $C$ is produced at the rate
$R(x,t)$ given by Eq.~(\ref{eq:ReakScale}) -- moves diffusively with its
centre at $x_f(t) = \sqrt{2 D_f t}$ and its width increasing as $w_f(t)
\sim t^{1/6}$.  Hence, the spinodal decomposition model can start with the
$C$ particles.  Their dynamics are described by a simple phase separating
equation taking particle conservation into account.  The specific dynamics
were introduced by Cahn and Hilliard \cite{Cahn1958,Cahn1961}.  We note
that an equivalent approach is the so-called model B
in critical dynamics \cite{Hohenberg1977}.

\begin{figure}\begin{center}\includegraphics[width=9cm]{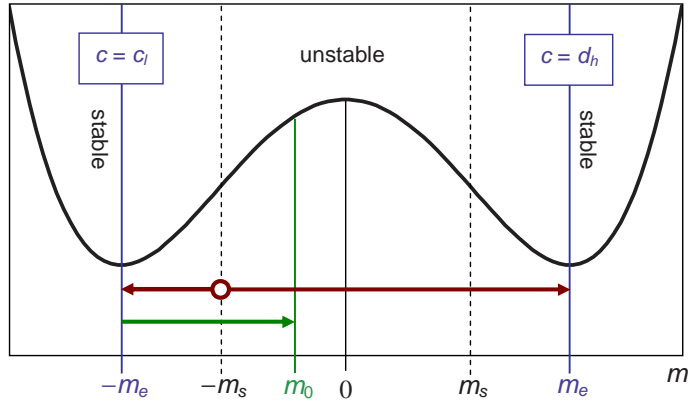}
\end{center}
\caption{(colour online) Illustration of the spinodal decomposition model. 
The homogeneous part of the free energy $F$ is shown as a function of the 
`magnetization' $m=c-(c_l+d_h)/2$.  See text for detailed explanation.
\label{fig:Fm}} \end{figure}

In the Cahn-Hilliard equation the concentration of $C$, $c$, is
represented by the so-called 'magnetization'
\begin{equation} m = c - (c_l+d_h)/2. \label{eg:magnet} \end{equation}
 Then a Ginzburg-Landau-type free energy is
defined in the simplest possible way required for obtaining two minima.
This free energy, illustrated in Fig.~\ref{fig:Fm}, is finally inserted into the
Cahn-Hilliard equation \cite{Racz1999},
\begin{equation} \partial_t m = - \lambda \partial^2_x \left[ \epsilon m
- \gamma m^3 + \sigma \partial^2_x m \right] + R(x,t), \label{eq:mart}
\end{equation}
Here, $R(x,t)$ is the source term introducing new $C$ particles into the
system via the $A+B \to C$ reaction.  The parameters $\epsilon$ and $\gamma$
have to satisfy $\sqrt{\epsilon/\gamma}=(d_h-c_l)/2$, while $\sigma$ must be
positive to eliminate short-wavelength instabilities.  $\lambda$ and $\sigma$
can be used to set the time-scale and length-scale of the system, which
leaves $\epsilon$ as the only free parameter.  Since $\epsilon$ measures the
negative deviation of the temperature $T$ from the critical temperature $T_c$,
$\epsilon>0$ is needed for $T<T_c$. No phase separation occurs for $T>T_c$.

The model can be understood most easily by looking at the free energy sketched
in Fig.~\ref{fig:Fm}.  The source term $R(x,t)$ moves the system form $m=-m_e$
(for $c = c_l \approx 0$) over the spinodal point $-m_s$ to $m=m_0$ which corresponds
to $c = c_0$.  Choosing $m_0$ such that it is located in the unstable regime, the
system will move to the second stable regime at $m=m_e$ corresponding to the high
concentration $c=d_h$, i.~e., band formation.  The band becomes a sink for $C$,
and its width grows until the front moves away so that $m$ can decay below
$m_s$, and the unstable regime is reached again, moving the system back to
$m=-m_e$.  Using this model it is possible to produce Liesegang patterns
satisfying the spacing law (\ref{eq:spacelaw2}) in agreement with the Matalon-%
Packter law (\ref{eq:MatalonPackterLaw}) \cite{Racz1999,Antal1999,Droz2000}.
Due to the conservation of $C$ particles the arguments regarding the width
law presented in Sect.~\ref{sec:widthlaw} also apply here.  Actually, the
derivation of the width law in Eq.~(\ref{eq:wn}) was introduced in the context
of the spinodal decomposition model.

A special feature not observed in the threshold models is the low density
phase with non-zero $c$, which has been reported in many experiments.  The
spinodal decomposition model can also be implemented for sophisticated
conditions.  For example, the effect of an additional electric field was
discussed recently this way \cite{Bena2005b}.  However, this can be done
similarly using a threshold model \cite{Bena2005}.

\subsection{The kinetic Ising model with Glauber and Kawasaki dynamics}\label{sec:kinIsing}

An alternative way to model the phase separating dynamics was recently
proposed by Magnin {\it et~al.} \cite{Antal2001} along the lines of the kinetic
Ising model for ferromagnets.  The main advantage is that this model fully
describes the fluctuations, going beyond the mean-field approximations.
Like in the spinodal decomposition model (see previous section), the
concentration of $C$ (and $D$) particles is represented as a magnetization.
The particles are identified as spin-up sites in a cubic lattice, while
spin-down sites represent vacancies.  Then the formation of precipitates
is modelled by a combination of spin-flip and spin-exchange dynamics
\cite{Droz1989}.  The Hamiltonian is the usual nearest neighbour Ising
Hamiltonian with ferromagnetic coupling $J>0$ between the spins
$\sigma_{\vec{r}}$ at site $\vec{r}$, modelling the attraction of the
$C$ particles,
\begin{equation} H=-J \sum_{{\rm neighbours} \, \vec{r}, \vec{r'}}
\sigma_{\vec{r}} \sigma_{\vec{r}'}. \end{equation}

Specifically, Glauber dynamics \cite{Glauber1963} are used to add the
$C$ particles in an initial state, in which all spins are down.  The
spin-flip rate $w_{\vec{r}}$ at site $\vec{r}$ is given by $w_r=R(\vec r,t)
(1-\sigma_{\vec{r}})/2$, with $R$ the source term discussed in 
Sect.~\ref{sec:ABC} (see Eq.~(\ref{eq:ReakScale})).  Since the width $w_f(t)
\sim t^{1/6}$ of the reaction front is not changing much, a constant width
leaving behind a constant concentration $c_0$ of $C$ particles was chosen
\cite{Antal2001}.  The diffusion and interaction of the spin-up sites,
i.~e., of the particles, is modelled by a spin-exchange process with
Kawasaki dynamics \cite{Kawasaki1966}, $w_{\vec{r} \to \vec{r}'}
= \left[ 1+\exp\left(\delta E/(k_B T) \right) \right]^{-1} /\tau_e$.
This exchange rate satisfies a detailed balance at temperature $T$.
The flip frequency $\tau_e$ sets the time scale, $T$ is the temperature,
$k_B$ the Boltzmann constant, and $\delta E$ the energy change.

2d and 3d simulations have been reported for this model
\cite{Antal2001}.  In 2d no pattern formation could be found
for a wide range of parameters, in contradiction with previous work using
lattice-gas simulations \cite{Chopard1994,Chopard1994b}.  In 3d 
patterns emerge in a restricted parameter range.  Due to limited
computational power no quantitative results for Liesegang patterns have
been published yet and fluctuations have not been analysed, indicating
that the model needs further investigation.

In the next section, we will thoroughly discuss lattice-gas simulations
for Liesegang pattern formation, since these studies have already yielded
quantitative results and fully include fluctuations.

\section{Lattice-gas simulations}\label{sec:latticegas}

In the previous section, we have seen that several mean-field models
can reproduce the basic laws of Liesegang pattern formation.  However,
there are problems not solved by mean-field models.
\begin{enumerate}
\item {\it Is the spacing law valid for all distances or only
asymptotically?}\\ While simple mean-field theories yield $x_n \propto (1+p)^n$
or equivalently $x_{n+1} = (1+p) x_n$, see Sect.~\ref{sec:Ionproduct},
experimental works usually report an asymptotic behaviour $x_{n+1}
\to (1+p) x_n$ for large $n$ (Eq.~(\ref{eq:spacelaw2})).
\item {\it Which version of the width law is the true one?}\\  While
  the mean-field theories generally yield $w_n \propto
  x_n^{\alpha}$ (see
Eq.~(\ref{eq:wn})), experiments have been fitted successfully by both
version of Eq.~(\ref{eq:widthlaw}).
\item {\it Which deviations from the Matalon-Packter law
(\ref{eq:MatalonPackterLaw}) or its more general mean-field form
(\ref{eq:pMF}) are relevant?}\\  Although most experimental findings are
consistent with the Matalon-Packter law (\ref{eq:MatalonPackterLaw}), one can
expect deviations if $D_B$ is large or if fluctuations become important.
\item {\it How stable are Liesegang patterns forming under different
conditions and in nanoscale systems?}\\ Simple mean-field models assume the 
existance of quasi-periodic Liesegang patterns and thus cannot be used to study 
the stability of the pattern formation process.  In general, mean-field quantities 
like concentrations are not well suited for studying nano-sized systems, since the 
number of atoms of one agent in a given small volume may fluctuate significantly.  
These fluctuations can increase or decrease the stability of the Liesegang patterns.
\item {\it What Liesegang patterns can be expected under special, restricted
(e.~g., dimensionally reduced), or designed geometries?}\\  This question is
particularly important if self-organization of Liesegang patterns shall be
applied to design nanoscale devices.  However, mean-field models usually
assume a quasi-1d geometry.
\end{enumerate}
In this section we will address problems 1 to 3 by means of lattice-gas simulations.
Problems 4 and 5 will be studied in later publications; we just briefly comment on 
them here.

Problem 4 can be addressed only by models that include the fluctuations of
the particle numbers and thus go beyond the mean-field limit.
Two major approaches in this direction have been published so far:
the kinetic Ising spin simulation reviewed in Sect.~\ref{sec:kinIsing} \cite{Antal2001},
and lattice-gas simulations on the basis of the nucleation and growth model
by Chopard {\it et~al.} \cite{Chopard1994,Chopard1994b}.  Since the Ising model
approach is still not so far advanced, we chose the second approach here.  An
additional advantage is the existence of a corresponding mean-field model
(see Sect.~\ref{sec:nucleationgrowth}), that the results can be compared with 
to find out deviations
in the universal laws and effects of fluctuations.  Equations~(\ref{eq:pMF}) and
(\ref{eq:wn}) represent the mean-field solutions for the spacing law and width
law, respectively.

Problem 5 has already been addressed by Liesegang himself since
the first experiments were in polar geometry. However, most theoretical work
was done in quasi-1d geometry, and nobody investigated how
geometry affects the pattern formation and the empirical laws. A first
step towards new geometries was done in \cite{Bensemann2005} using mean-field
models plus a stochastic term. However, this ansatz seems to be hard to
control for some special geometries. Therefore we think that the lattice-gas
model can serve as a good candidate to test also new geometries and their
influence on the pattern formation.

Lattice-gas simulations can serve as a computational
experiment to test how the mean-field solutions can be applied to `experimental'
data and furthermore how fluctuations might cause deviations from these
solutions. As suggested by the separation of the two processes in the nucleation and
growth model, the lattice-gas simulation consists of two stages.  In a first
stage the $C$ particles are generated (cf. Sect.~\ref{sec:ABC}).  The second
stage simulates the precipitation of the $C$ particles by rules comparable
with cellular automata (cf. Sect.~\ref{sec:CtoD}).

\subsection{First process: reaction-diffusion $A+B \to C$}
\label{sec:LGABC}

In the simulation, we consider a simple cubic lattice of size $M \times M
\times L$ with $L \gg M$, see Fig.~2(a).  Each lattice site represents one 
cubicle (cell) of the system.  Initially, the lattice is homogeneously filled 
with $B$ particles, i.~e., there are on average $b_0$ independent $B$ particles
on each site.  The $A$ particles are placed on the left plane of the lattice
with $a_0$ particles on each of the $M \times M$ sites.  The parameters $a_0$ 
and $b_0$ can thus be interpreted as concentrations per lattice site.  Typically,
$b_0$ is in between 10 and 200, while $a_0$ is about twice as large.  The
mean-field limit can be reached if either the considered cells are
enlarged or the number of particles per cell is increased.  Thus, increasing
both $a_0$ and $b_0$ but keeping their ratio constant, drives the simulation
to the mean-field limit.

The dynamics is modelled as follows.  To simulate diffusion, both, $A$
and $B$ particles perform independent random walks on the lattice.  The
diffusities $D_A$ and $D_B$ are defined as the probabilities that a motion
in either of the six possible directions takes place in a given time step
\cite{Chopard1991,Kkarapiperis1994}, i.~e., $D_A$ and $D_B$ are proportional
to the physical diffusivities.  After each time step, the left plane
of the lattice is re-filled with $A$ particles with the initial condition.
A reaction $A+B\to C$ takes place with probability $\kappa$ ($\kappa = 1$
in our simulations), if at least one $A$ and one $B$ particle is found on
the same site \cite{Cornell1991,Chopard1991b}.  Afterwards, all $C$
particles also diffuse independently on the lattice with diffusity $D_C$.
It was previously shown that the mean-field predictions regarding the
reaction rate $R(x,t)$ (reviewed in Sect.~\ref{sec:ABC}) are in agreement
with the simulations in 3d \cite{Cornell1991}, while
logarithmic corrections occur in 2d \cite{Chopard1991b}.

\begin{figure}
\begin{center} \includegraphics[width=13cm]{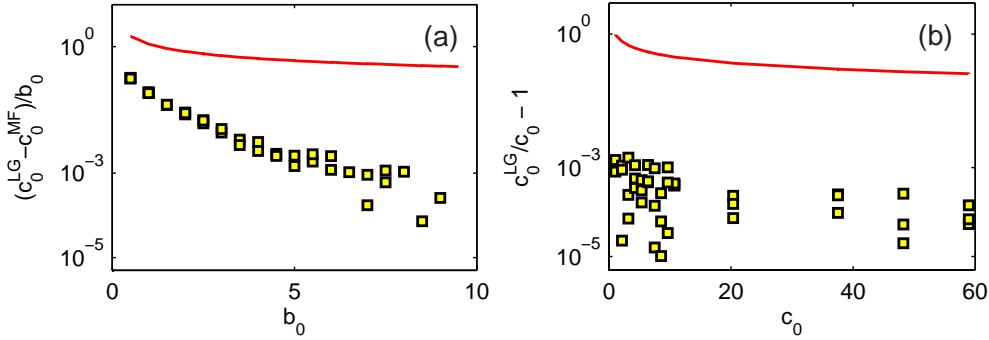}\end{center}
\caption{(colour online) (a) Deviations of the 3d lattice-gas simulations from
the mean-field limit. The values of simulated ratios $c_0^{\rm LG}/b_0$ have been
calculated from the plateau of $c(x,t)$ observed behind the diffusion front
(see Fig.~\ref{fig:nucgrowth}(b)).  The corresponding mean-field value
$c_0^{\rm MF}/b_0$ was calculated from Eq.~(\ref{eq:c0}). The red curve marks
the (scaled) standard deviations of the particle numbers per lattice cell from
their means $c_0^{\rm LG}$ indicating the fluctuations in the production of the
$C$ particles. The simulation parameters are $a_0/b_0=2$, $D_A=1$, $D_B=0.1$.
(b) Results of our lattice-gas simulations with $C$ particles inserted in an
approximated reaction zone according to Eqs.~(\ref{eq:Rxt}) and (\ref{eq:A}).
The values of simulated ratios $c_0^{\rm LG}/c_0$ have been calculated as in (a).
They are in good agreement with the inserted $c_0$ for all concentrations.  Red
curve same as for (a). The parameters are $D_f=1.22$ (calculated from Eq.
(\ref{eq:Dfzwei})) and $\Delta=3$.
\label{fig:ABtoC}}
\end{figure}

Here, we use for the first time 3d lattice-gas simulations for Liesegang
patterns in contrast to the 2d setup employed by Chopard {\it et~al.}
\cite{Chopard1994}. For an alternative 3d simulation see Sect.~\ref{sec:kinIsing} 
\cite{Antal2001}. To confirm that the mean-field limit is reached for
large concentrations, we ran simulations with different $b_0$ (keeping
$a_0/b_0$ constant) and found that $c_0/b_0$ becomes constant in agreement
with Eqs.~(\ref{eq:Dfeins}) and (\ref{eq:c0}).  Figure~\ref{fig:ABtoC}(a) shows
the simulated deviations from the mean-field limit, which decay below one
percent for $b_0 \gg 2$. We focus on $b_0 > 10$, where the deviations are
below $10^{-4}$.  In a 2d (or 1d) simulation, fluctuations would alter $c_0$
such that $c_0/b_0$ does not converge to the mean-field limit even for large
$b_0$ and $a_0$ \cite{Chopard1991b}.

Since the mean-field solutions are valid for a 3d setup, it is possible to
insert the $C$ particles directly.  This ansatz was first used in a kinetic
Ising scenario \cite{Antal2001} (see Sect.~\ref{sec:kinIsing}).  We employ this idea to
speed up our numerical calculations, since fully simulating the reaction-diffusion
process $A+B\to C$ reduces the computational speed by at least one order of
magnitude. Otherwise we could not work with concentrations $b_0>10$ in sensible
time.  However, this approximation reduces the fluctuations of $C$ particle
production, in particular for small concentrations.

The probability to insert $C$ particles into the lattice is given by the
reaction rate $R(x,t)$ (see Eq.~(\ref{eq:ReakScale})), with a centre position
$x_f(t)$ moving as $x_f(t) = \sqrt{2 D_f t}$ to the right.  Usually, $D_f$ is
calculated via Eq.~(\ref{eq:Dfeins}), assuming that the $A$ particles are
distributed homogeneously for $x \to -\infty$, i.~e., $a(-\infty,t) = a_{-\infty}$.
In this case the $A$ particle concentration outside the quasi-stationary region
(see Sect.~\ref{sec:ABC}) can be approximated by $a(x,t) = a_{-\infty} - a_f
\left[\erf\left(\frac{x}{2 \sqrt{D_A \, t}}\right) +1 \right]$ with a constant $a_f$.
However, in our configuration, Eq.~(\ref{eq:Dfeins}) needs to be modified, since
the concentration of $A$ particles is held constant for all times on the left plane,
i.~e., for $x=0$:  $a(0,t) = a_{-\infty} - a_f = a_0$. A straightforward derivation
similar to the derivation of Eq.~(\ref{eq:Dfeins}) in \cite{Koza1996} leads to
the solution
\begin{equation}
  \erf\left(\sqrt{\frac{D_f}{2 D_A}}\right) \exp\left(\frac{D_f}{2 D_A}\right) =
  \frac{a_0 \sqrt{D_A}}{b_0 \sqrt{D_B}} \, H\left[\sqrt{\frac{D_f}{2
  D_B}}\right].
  \label{eq:Dfzwei}
\end{equation}
Following \cite{Antal2001} we approximate the nearly Gaussian-shaped reaction rate
term $R(x,t)$ by
\begin{equation}
  \tilde R(x,t) = \frac{\tilde A_f}{\sqrt{t}} \Theta(x-x_f+\Delta)
  \Theta(x_f + \Delta -x),  \label{eq:Rxt}
\end{equation}
with constant width $2\Delta$ of the front.  These simplifications are justified
since the band forming process does not depend on the exact form of the reaction
zone \cite{Antal1999} and the width of the front increases very slowly in time
(as $t^{1/6}$, see Eq.~(\ref{eq:ReakScale})).  The most important feature of the
reaction front is that it leaves behind a constant concentration $c_0$ of $C$
particles.  Therefore, $\tilde A_f$ in Eq.~(\ref{eq:Rxt}) is chosen to be
\begin{equation}
  \tilde A_f= \frac{\sqrt{2 D_f}}{4 \Delta} c_0.
  \label{eq:A}
\end{equation}
Figure~\ref{fig:ABtoC}(b) compares the simulated $c_0^{\rm LG}$ with the $c_0$ value
inserted into the simulation via Eq.~(\ref{eq:A}). In contrast to a full simulation
of the $A+B \to C$ reaction (see Fig.~\ref{fig:ABtoC}(a)), $c_0^{\rm LG}$ reaches
the mean-field value already for small concentrations since the fluctuations induced
by the motion of the $A$ and $B$ particles are eliminated.  Only the standard
deviation (red curve) of the number of particles per cell is similar as in 
Fig.~\ref{fig:ABtoC}(a), since the reaction front in (b) is an approximation of the
reaction front in (a) and both cause the same fluctuations in $C$ particle production.

\subsection{Second process: precipitation $C \to D$}
\label{sec:CtoD2}

The second part of our lattice-gas simulation is the precipitation of $C$
particles, generating the immobile precipitate $D$.  The nucleation of $D$
depends only on $c(x,t)$, while the growth of $D$ depends on both, $c(x,t)$
and $d(x,t)$, see also Sect.~\ref{sec:CtoD}.  In the particle picture of
our lattice-gas simulation the density will be the number of particles
divided by the considered cell volume. Here, this considered volume around
a given lattice cell includes all neighbour cells, i.~e., the 27 cells in 
a cube of $3 \times 3 \times 3$ cells.  This definition sets the mean 
reaction distance.

Two thresholds are introduced. If the mean local concentration of $C$
particles (in the 27 cells) exceeds a threshold $K_{sp}$, nucleation
occurs.  $C$ particles in the vicinity of $D$ particles precipitate already
if their mean local concentration exceeds a threshold $K_p < K_{sp}$.
$C$ particles on top of $D$ particles always precipitate, which makes
the growth process fast enough to deplete a region of $C$ particles.
Without this option the growth process would not terminate, and no
additional Liesegang bands could be formed.  The resulting bands have
distinct positions, which we will denote by $x'_n$ in the following;
their width is denoted by $w_n$.  We calculated $x'_n$ as the centre
of the bands, i.~e., the mean of the positions of the first and last
$D$ particles within each band.  The corresponding widths $w_n$ are
defined as the differences between these two positions.

\subsection{Reconsideration of the spacing law} \label{sec:reconspacing}

Equation~(\ref{eq:spacelaw2}), i.~e., the asymptotical convergence of the
ratio of the positions of neighbouring bands $x_{n+1}/x_{n} \to 1+p$ for
large $n$, was estimated from empirical findings; see Sect.~\ref{sec:basiclaws}.
On the other hand, theoretical analyses suggest that the ratio $x_{n+1}/x_{n}$
is identical with $1+p$ \cite{Wagner1950,Prager1956,Zeldovich1961,Keller1981},
as was discussed in Sects.~\ref{sec:Ionproduct} and \ref{sec:nucleationgrowth}.
Although these two forms of the spacing law contradict each other, both are
still used in parallel without much considerations. We propose a way to
reconcile the experimental findings described by the empirical form of the
spacing law with the theoretical mean-field prediction.

\begin{figure}
\begin{center}\includegraphics[width=7cm]{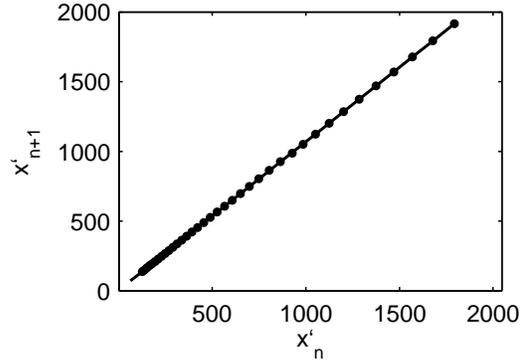}\end{center}
\caption{Spacing law for simulated data depicted in a way different from 
Fig.~\ref{fig:BspLS}(b), where $x'_{n+1}/x'_n$ was plotted versus $n$ to 
observe the asymptotical behaviour (cf. Eq.~(\ref{eq:spacelawsim})). Here, 
$x'_{n+1}$ is plotted versus $x'_n$ to observe the linear behaviour with 
slope $1+p$ and offset $p\xi$ (cf. Eq.~(\ref{eq:spacelawsim2})), yielding 
$p=0.066$ and $\xi=62$.  Parameter set for the 3d lattice-gas simulation: 
$a_0=130$, $b_0=65$, $D_A=1$, $D_B=0.1$ (leading to $D_f=1.22$ and 
$c_0/b_0=1.072$); $D_C=0.1$, $K_{sp}/b_0=0.93$, and 
$K_p/b_0=0.52$. 
\label{fig:MCspacelaw}}
\end{figure}

In experimental and numerical results the position of the first Liesegang band
is usually blurred and thus not well defined, contrary to theoretical analyses.
Thus, there is an arbitrariness in choosing the position of the first band.
Therefore, it is better to use different variables $x'_n$ for the experimentally
or numerically measured band positions and $x_n$ for the theoretical positions,
since there may be an offset $\xi$ between them,
\begin{equation}
x_n = x'_n + \xi , \label{eq:defb}
\end{equation}
due to, e.~g., the blurred first band.  Another possibility is that the mass of $D$ 
in the first Liesegang band as well as the band's width cannot be close to zero 
as would be required if the linear increase observed in Fig.~\ref{fig:nucgrowth}(c) 
was beginning at $x'=0$.  We will see later that usually $\xi > 0$, i.~e., $x_n > 
x'_n$, indicating that the ideal laws are based on an `imaginary' starting point 
outside the real sample.  In literature, however, significant confusion is 
caused by the fact that both variables, $x_n$ and $x'_n$, are not distinguished.

Studying measured values $x'_n$, and assuming both, $x'_n = x_n - \xi $ and the
ideal mean-field result, $x_{n+1} = (1+p) x_{n}$, one finds the asymptotic
(empirical) form of the spacing law,
\begin{equation}
\frac{x'_{n+1}}{x'_n} = {x_{n+1} - \xi  \over x_n - \xi } 
= 1 + p {x_n \over x_n - \xi } = 1+ p {x'_n +\xi  \over x'_n} = 1 + p +
\frac{p\xi }{x'_n} \to (1+p) \label{eq:spacelawsim}
\end{equation}
for large $n$ as in Eq.~(\ref{eq:spacelaw2}).  Figure~\ref{fig:BspLS}(b)
shows this asymptotic behaviour as observed in simulated data; note that 
actually the values of $x'_n$ rather than $x_n$ are plotted.  The coefficient 
$p$ is hard to determine in such plots, especially if only few bands
are present.  A more convenient way to extract $p$ from measured data is to fit
the equation
\begin{equation}
 x'_{n+1}=(1+p) x_n' + p\xi, \label{eq:spacelawsim2}
\end{equation}
where $p\xi$ is the intersection with
the ordinate, as shown in Fig.~\ref{fig:MCspacelaw} for the same data.
Even for a small number of bands, $p$ and $\xi$ can be extracted conveniently.
In addition, we believe that the offset $\xi$ is an important ingredient that
should not be neglected in interpreting any experimental or simulated data.
However, it remains unclear if or how the value of $\xi$ could be predicted.

\subsection{Reconsideration of the Matalon-Packter law} \label{sec:reconMP}

\begin{figure}
\begin{center} \includegraphics[width=13cm]{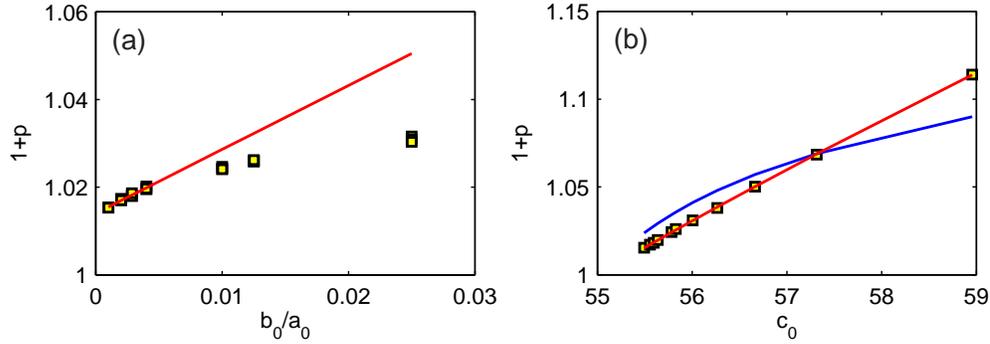}\end{center}
\caption{(colour online) Reconsideration of the Matalon-Packter law based on 3d lattice-gas 
simulations. (a) The yellow squares indicate simulation results for the spacing law 
coefficient $p$ obtained keeping $b_0$ constant
and varying $a_0$.  The straight line is a linear fit to the first four data points
using Eq.~(\ref{eq:MatalonPackterLaw}).  (b) Simulation results similar to those shown in 
(a). $D_f$ and $c_0$ are calculated using Eqs.~(\ref{eq:Dfzwei}) and (\ref{eq:c0}), 
respectively. The blue 
curve is a fit of Eq.~(\ref{eq:pMF}), while the red curve takes fluctuations into account.
Parameters: $b_0=55$, $D_A=1$, $D_B=0.1$, $D_C=0.15$, $K_{sp}/b_0=0.96$,
and $K_p/b_0=0.52$. 
\label{fig:MCp}}
\end{figure}

The original Matalon-Packter law (\ref{eq:MatalonPackterLaw}) predicts a linear
dependence between $b_0/a_0$ and $p$.  It was shown in Sect.~\ref{sec:MPlaw} that
this empirical law holds for $b_0 \ll a_0$ only.  In addition, a more general law
was presented in Eq.~(\ref{eq:pMF}) \cite{Antal1998}.  The deviations from
Eq.~(\ref{eq:MatalonPackterLaw}) are even stronger in the results of lattice-gas
simulations.  An example is depicted in Fig.~\ref{fig:MCp}(a).  A linear
dependence is observed only for $a_0/b_0 > 250$, a criterion shown to be equivalent
to $D_B/D_f \approx 1$ \cite{Antal1998}.  If $a_0$ is reduced for constant $b_0$,
$D_f$ increases, driving the system out of the regime in which Eq.
(\ref{eq:MatalonPackterLaw}) is valid \cite{Antal1998}. Our lattice-gas
simulations confirms these results, see Fig.~\ref{fig:MCp}(a).

Figure~\ref{fig:MCp}(b) compares a fit of the generalized Matalon-Packter law
Eq.~(\ref{eq:pMF}) (blue curve) with the results of our lattice-gas simulations.
The deviations are still quite large. Only if fluctuations are taken into account
by further modifying the generalized Matalon-Packter law, a nice agreement
can be reached.  The formula used for the red fit in Fig.~\ref{fig:MCp}(b)
will be discussed and motivated in detail in a later publication.

\subsection{Reconsideration of the width law} \label{sec:reconwidth}

\begin{figure}
\begin{center} \includegraphics[width=13cm]{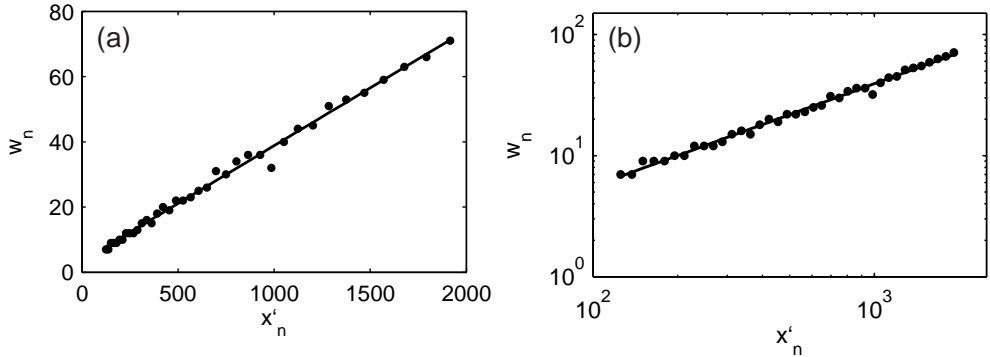}\end{center}
\caption{Reconsideration of the width law based on 3d lattice-gas simulations.
(a) Linear presentation of $w_n$ versus $x'_n$ (points) together with a fit of the first 
version of Eq.~(\ref{eq:widthlaw})
with $\mu_1=0.035$ and $\mu_2=3.4$. (b) Double logarithmic presentation of the same data
together with a fit of the second version of Eq.~(\ref{eq:widthlaw}) with slope
$\alpha=0.82$.  The same simulation as for Figs.~\ref{fig:BspLS} and \ref{fig:MCspacelaw} 
was used. \label{fig:BspWL}}
\end{figure}

As discussed in Sect.~\ref{sec:basiclaws} two competing empirical forms of the width
law are used for fitting experimental data,
$$ w_n = \mu_1 x_n + \mu_2 \qquad {\rm and} \qquad w_n \propto
x^{\alpha}_n. \eqno{(\ref{eq:widthlaw})} $$
Figure~\ref{fig:BspWL} shows the results of our lattice-gas simulations in both
representations, because $w_n$ is plotted versus $x'_n$ both linearly and double
logarithmically.  Clearly, it is not possible to favour either of the two forms of the
width law, since all widths are small compared with the size of the sample and both
fits have an equivalent quality.

On the other hand, in Sect.~\ref{sec:widthlaw} a simple and general width law was
deduced theoretically (see Eq.~(\ref{eq:wn})) using particle conservation.  Assuming
scaling behaviour of the band densities $d_n$, $d_n \propto x_n^\beta$, and setting
$c_l = 0$ (since no precipitate occurs between the bands in our simulation), one obtains
\begin{equation}
w_n = p{c_0 \over d_n} x_n \propto x_n^{\alpha} \label{eq:wn2}
\end{equation}
with $\alpha = 1-\beta$.  This law was derived using the theoretical $x_n$ where
the ratio for two consecutive bands is constant, i.~e., the exact form of the spacing
law.  A width law valid for experimental or numerical data of the positions $x'_n$
of the bands can be derived introducing the offset $\xi $, see Eq.~(\ref{eq:defb}).
The analytical form of the width law thus becomes
\begin{equation}
w_n = p \frac{c_0}{d_n} (x'_n + \xi ) = p \frac{c_0}{d_n} x'_n + \frac{c_0}{d_n} p \xi ,
\label{eq:wnaus}
\end{equation}
which is in between the two competing empirical forms (\ref{eq:widthlaw}), since
$d_n \propto (x'_n+\xi )^\beta$.  Note that $p \xi $ is identical with the offset of the
fit for the spacing law (\ref{eq:spacelawsim2}).  Hence, the parameter $\xi $ is also
important for understanding the width law.

\begin{figure}
\begin{center} \includegraphics[width=13cm]{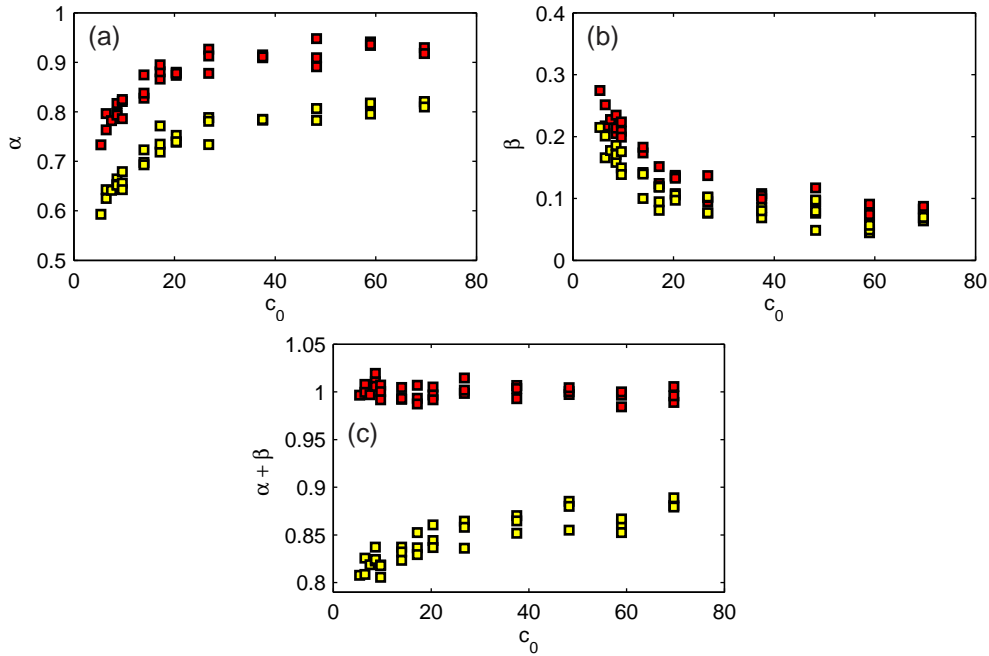}\end{center}
\caption{(colour online) Fits of (a) width, (b) density, and (c) mass scaling behaviour based on our
3d lattice-gas simulations for different values of $c_0$. The exponents have been
obtained by fitting a power-law to plots of (a) $w_n$, (b) $d_n$, and (c) $m_n=d_n w_n$
versus $x'_n$ (yellow dots) and versus $x'_n+\xi $ (red dots).  The parameters of the
simulations are identical with those used for Figs.~\ref{fig:BspLS} and \ref{fig:MCspacelaw}, 
except for varied $c_0$.\label{fig:MCwidthlaw}}
\end{figure}

To test these predictions we have run simulations and fitted Eq.~(\ref{eq:wn2}) to
extract $\alpha$ from the data disregarding $\xi $, i.~e., inserting $x'_n$ directly
for $x_n$.  The results are shown in Fig.~\ref{fig:MCwidthlaw}(a), yellow points.
In a second approach we used the same data and inserted $x_n = x'_n+\xi $ into Eq.
(\ref{eq:wn2}), also obtaining $\alpha$ as a fit parameter.  The results are shown
in Fig.~\ref{fig:MCwidthlaw}(a), red points.  Both approaches lead to different
values of $\alpha$ indicating that the offset $\xi $ plays a significant role. 

Since the width law depends on the density of the bands $d_n$, it is not possible to
distinguish the correct form by just looking at the width exponent $\alpha$. Therefore,
we have also calculated the densities $d_n$ in the numerical simulations.  
Figure~\ref{fig:MCwidthlaw}(b) shows the exponent $\beta$ obtained by fitting $d_n \propto
{x'_n}^\beta$ (yellow points) and $d_n \propto (x'_n+\xi )^\beta$ (red points).  In addition,
Fig.~\ref{fig:MCwidthlaw}(c) depicts the sum of both exponents from parts (a) and (b),
$\alpha + \beta$.  The product $w_n d_n$ can be interpreted as the mass $m_n$ of band
$n$; it scales as $m_n = p c_0 x_n = p c_0 x'_n + p c_0 \xi $ according to Eq.~(\ref{eq:wnaus}).  
Thus, according to theory, the mass exponent $\alpha + \beta$ must
be one.  The results shown in Fig.~\ref{fig:MCwidthlaw}(c)
indicate that this holds only
in the case where the offset $\xi $ is taken into account correctly (red points),
confirming that $\xi $ must not be disregarded.

\subsection{Application to previous experimental data}

We believe that much confusion in literature could be avoided if the difference between
$x_n$ and $x'_n$ was taken into account. In particular, the long-standing discussion
about the width law could probably be solved if the analysis of experimental data
included plots versus $x_n=x'_n + \xi $.  We suggest that former experiments should be 
re-analysed using $\xi$ and $x'_n$ to confirm our conclusions.

An experiment which confirms the result of Eq.~(\ref{eq:wnaus}) was published
recently \cite{Narita2006}. The authors investigated pattern formation in a
$\kappa$-Carrageenan gel.  They measured $x'_n$ and $w_n$ for different concentration
of the outer solution.  First they plotted $x'_{n+1}$ versus $x'_n$ in a fashion
similar to Fig.~\ref{fig:MCspacelaw} and extracted $p$.  Looking at the plot,
we can clearly see offsets corresponding to $p\xi$.  Secondly they plotted $w_n$ versus
$x'_n$ in a linear fashion similar to Fig.~\ref{fig:BspWL}(a), extracting the
parameter $\mu_1= p c_0 / d_n$ in Eq.~(\ref{eq:wnaus}), assuming constant densities
$d_n$.  Looking at the plot, we also clearly see offsets corresponding to $\mu_2 = p \xi 
c_0 /d_n$ (according to Eq.~(\ref{eq:wnaus})).  In addition they determined $c_0/d_n=
0.53$.  This means that the offsets of the width law must be approximately half of
the offsets of the spacing law.  This conclusion is in quantitative agreement with
the offsets we read from the plots in \cite{Narita2006}.  The experimental data thus 
confirm our conclusion that the offset $\xi$ should be taken into account.

The differences of Eqs.~(\ref{eq:wn}) and (\ref{eq:wnaus}) explain the different
approaches used in experiment and theory.  In summary, we have
solved problems 1 and 2 raised in the beginning of
Sect.~\ref{sec:latticegas}; problem 3 was discussed in
Sect.~\ref{sec:reconMP}.

\section{Summary and outlook}\label{sec:sumoutlook}

In this paper we reviewed the four empirical laws believed to govern Liesegang
pattern formation and several important mean-field models reproducing these laws
as well as a few Monte-Carlo-type simulations.  Based on our detailed 3d lattice-gas
simulations of the nucleation and growth model, we detected three major problems
in reconciling the experimental reports with mean-field results.  However, our
simulations also helped us to find a straightforward solution.

We have shown that all basic empirical laws describing Liesegang pattern formation,
i.~e., the time law, the spacing law, the Matalon-Packter law, and the width law can
be understood on the basis of the nucleation and growth model.  A full agreement
between experimental observations, simulation results, and mean-field models can be
obtained only if a constant offset $\xi$ between measured and theoretically assumed
band positions is taken into account, as suggested in this work.  A re-analysis of a
previous experiment concerning the spacing law and the width law confirmed our
suggestion.  We propose that further experiments should be re-analysed to test the
refined predictions.

In addition, we suggest further experiments with systematically varied concentrations
of both, inner and outer electrolyte to confirm or extend the generalized
Matalon-Packter law (\ref{eq:pMF}).  Our current work in progress regarding the
effects of fluctuations on Liesegang pattern formation indicates that additional
modifications of the Matalon-Packter law are necessary in small-scale systems.
Reliable experimental results for the nanoscale regime, where fluctuations are
definitely important, are expected to become available soon, since experiments with
nanoscale particles get feasible.  A big open question will be how to manipulate the
pattern formation such that interesting nanoscale devices could be designed.  A
deeper understanding of the dynamics is important for reaching this goal.
For a first promising work in this direction, see \cite{Antal2007}.

\begin{acknowledgement}\label{sec:acknowledgement}
We thank I. Bena, M. Droz, M. Dubiel, I. L'Heureux, Y. Kaganovskii, I. Lagzi,
K. Martens, and M. Rosenbluh for discussions. We would like to acknowledge
support of this work from the Deutsche Forschungsgemeinschaft (DFG,
project B16 in SFB 418).
\end{acknowledgement}


\begin{thebibliography}{10}
\bibitem{foot:MPlaw}
In the original work, the Matalon-Packter law was introduced as $p=F(b_0)
  + G(b_0)/a_0$. The disadvantage of this equivalent definition is that
  $G(b_0)$ gets the dimension of $b_0$.
\bibitem{foot:Zrinyi}
In the original paper the second term was not introduced as a formula but
  described in the text.
\bibitem{foot:a0}
For our problem we will need the additional boundary condition $a(0,t)=a_0$.
  The changes needed to achieve this will be introduced in Sect. 4.1.
\end{thebibliography}
\end{document}